\documentclass[journal=jacsat,manuscript=article]{achemso}
\usepackage[dvipsnames,table]{xcolor}
\usepackage{pdfpages}
\usepackage[version=3]{mhchem} %
\usepackage{miller}
\usepackage{subcaption}
\usepackage{multirow}
\usepackage{color}
\usepackage{textgreek}

\graphicspath{{Figures/}}
\usepackage{xspace}
\usepackage{comment}
\usepackage{graphicx}
\usepackage[%
  breaklinks,             %
  bookmarks = false, %
  pdfpagemode   = UseNone,%
  pdfstartview  = FitH,     %
  pdfstartpage  = 1,      %
  colorlinks    = true,   %
]{hyperref}

\newcommand\myshade{85}
\colorlet{mylinkcolor}{YellowOrange}
\colorlet{myurlcolor}{Aquamarine}
\colorlet{mycitecolor}{violet}

\hypersetup{
  linkcolor  = mylinkcolor!\myshade!black,
  citecolor  = mycitecolor!\myshade!black,
  urlcolor   = myurlcolor!\myshade!black,
  colorlinks = true,
}



\author{Abrar Fahim Navid}
\affiliation{Department of Mechanical Engineering, Texas Tech University, Lubbock, Texas 79409, USA}
\author{Zeeshan Ahmad}
\affiliation{Department of Mechanical Engineering, Texas Tech University, Lubbock, Texas 79409, USA}
\email{zeeahmad@ttu.edu}

\title{Understanding halide segregation in metal halide perovskites through defect thermodynamics}

\keywords{}

\begin{document}

\begin{abstract}

Halide segregation in metal halide perovskites limits their bandgap tunability and hinders their adoption in tandem solar cells and light emitting diodes. Here, we reveal the thermodynamic driving force behind halide segregation in mixed halide (Br-I) perovskites. By performing first-principles calculations on slab models with varying bromide and iodide distributions, we demonstrate that bromide ions preferentially occupy surface sites over bulk sites. Our simulations show that the segregation tendency is higher in MAPb(Br$_x$I$_{1-x}$)$_3$ (MA=methylammonium) compared to FA$_{0.8}$Cs$_{0.2}$Pb(Br$_x$I$_{1-x}$)$_3$, highlighting the role of the A-site cation. To quantify this effect, we establish a descriptor for halide segregation: the difference in defect formation energies of Br antisite defects between the bulk and the surface, which confirms the role of the A-site cation at equimolar Br-I concentration. Furthermore, we identify the localization of photo-generated holes near iodide ions, which triggers their oxidation and accelerates the formation of iodide vacancies, thereby promoting segregation. Overall, this work establishes defect thermodynamics as a framework for understanding halide segregation and provides a structural basis for designing stable mixed halide perovskites.

\end{abstract}

\section{Introduction}

Hybrid organic-inorganic perovskites are leading candidates for next-generation solar cells due to high power conversion efficiencies exceeding 25\% and compatibility with low-cost and scalable solution processing techniques~\cite{Green2014emergence,zhangProgressIssuesPin2024,dasadhikariDecodingRecombinationDynamics2025,huangUnderstandingPhysicalProperties2017,jiangSurfaceReactionEfficient2022}.  The high efficiency stems from their superior optoelectronic properties such as long carrier lifetimes, defect tolerance, strong light absorption, and high carrier mobility. However, further improvements in efficiency and stability are required to compete with existing silicon solar cells for eventual commercialization~\cite{yangAchievementsChallengesFuture2024}. One route to commercialize perovskite solar cells is to develop perovskite-silicon tandem cells, which can circumvent the limits of single-junction solar cells~\cite{shockleyDetailedBalanceLimit1961}. This requires bandgap tunability of perovskites, which can be achieved by doping of the halide site, forming mixed halide perovskites~\cite{mcmeekinMixedcationLeadMixedhalide2016}. Bandgap tunability is also essential to improve the efficiency of single junction perovskite solar cells and the development of light-emitting diodes with tunable emission wavelengths\cite{busipalli-2019,Noh-mixedhalide,friend-mixedhalide}. The bandgap of methylammonium  lead iodide (\ce{MAPbI3}) can be increased by Br doping.
However, a major challenge in the design of mixed halide perovskites is the segregation of halides during illumination, resulting in the formation of I-rich and Br-rich phases. Notably, light-induced segregation in MAPb(Br$_{x}$I$_{1-x}$)$_3$ consistently produces I-rich domains near $x \approx 0.2$, suggesting a particularly stable compositional endpoint. \cite{brennanLightInducedAnionPhase2018}
This phenomenon significantly reduces the open circuit voltage and impacts the stability, efficiency, and operational performance of devices, particularly under continuous illumination \cite{Slotcavage-2016,datta,origin-voltageloss}. The iodide-rich domains have narrower bandgaps, which help recombine charges and increase radiative recombination rates\cite{draguta2017rationalizing,hoke2015reversible,Ruth, brennanLightInducedAnionPhase2018}. 

Understanding the mechanism and factors responsible for halide segregation is essential to develop strategies to stabilize mixed halide perovskite solar cells. 
Existing studies suggest  that halide segregation is driven by halide defects~\cite{yoonShiftHappensHow2017}, carrier-generation gradients within the perovskite film~\cite{barkerDefectAssistedPhotoinducedHalide2017}, halide oxidation~\cite{KERNER20212273,photolysis}, hole trapping by iodine~\cite{Dubose-kamat-holetrapping}, iodine loss~\cite{leeEvidence2Loss2025}, and ion migration~\cite{kimLargeTunablePhotoeffect2018,senocrateSolidStateIonicsHybrid2019}. 
Under illumination, holes can displace lattice iodide to form iodine vacancies which can happen to a different degree in I and Br. The different tendency of I and Br ions to move towards the surface can drive phase segregation~\cite{senocrateSolidStateIonicsHybrid2019}. %
While halide segregation is generally reversible\cite{hoke2015reversible,halford-2022}, loss of \ce{I2} can cause irreversible structural changes~\cite{leeEvidence2Loss2025}.
Some strategies proposed for suppressing segregation include  A-site substitution,\cite{safdari-2024,kamruddin-2022,knigh&hertzt2020preventing,knightHalideSegregationMixedHalide2021, Huang-2024} lattice compression,\cite{Lattice-compression} using excess halides,\cite{yoonShiftHappensHow2017} controlling the ionic mobility,\cite{hoke2015reversible,halford-2022,KUNO202021,suppression-periodicheating} using self-assembled long chain organic ammonium capping layers,\cite{Xiao-prevent-segregation} using 2D perovskite architecture\cite{Hiott-2023,Cho-2020} etc. In all mechanisms, it is evident that defects such as I vacancies and Br antisites play a critical role since segregation involves the creation, annihilation, and rearrangement of these defects. However, most studies to date have focused on light-driven kinetics of the rearrangement process rather than the intrinsic thermodynamics of defects responsible for segregation in the mixed halide perovskite structure. Further, these defects behave differently near surfaces, interfaces, and grain boundaries, which can serve to initiate the segregation process \cite{ahmadUnderstandingEffectLead2022,mottiControllingCompetingPhotochemical2019,mottiPhaseSegregationMixedhalide2021,tangLocalObservationPhase2018}. 
For example, ~\citet{tangLocalObservationPhase2018} showed segregation preferentially happens near the grain boundaries rather than the grain centers. This shows a clear need for proper analysis of halide segregation behavior from bulk to surface.

In this work, we establish the thermodynamic origin of defect-mediated segregation using first-principles calculations of mixed halide perovskites with  different distributions of I and Br within the slabs.  We demonstrate that Br ions prefer to occupy sites near surfaces, establishing a compositional gradient from the surface to the bulk that follows an exponential profile.
The strength of this surface-directed segregation is dictated by the A-site cation and is most pronounced in MA and significantly reduced for the FA$_{0.8}$Cs$_{0.2}$ (FA=formamidinium) cation  in agreement with experiments~\cite{knightHalideSegregationMixedHalide2021,mcmeekinMixedcationLeadMixedhalide2016}. Our analysis further shows that this tendency for Br localization near the surface can be rationalized by analyzing the Br–Pb bond lengths and Pb-Br-Pb bond angles of the local coordination environment. Furthermore, probing the influence of light, we identify that the holes are localized near I sites, promoting their oxidation. This process leads to more iodide vacancy generation, providing sufficient ion conductivity for structural rearrangement.  Finally, our simulations of halide vacancies reveal that I vacancies are formed in preference to Br vacancies near surfaces and in the bulk, promoting the irreversible loss of \ce{I2} gas through the surface~\cite{leeEvidence2Loss2025}. 

\section{Methods}
First-principles density functional theory (DFT) calculations were performed using the Quantum ESPRESSO package\cite{giannozziQUANTUMESPRESSOModular2009,giannozziAdvancedCapabilitiesMaterials2017}. The cubic phase was used for both \ce{MAPbI3}
and \ce{FAPbI3} in this study to remain consistent with the experimental observations reported in the literature on halide segregation\cite{knightHalideSegregationMixedHalide2021}. The lattice parameters of 6.276 \AA\  and 6.362 \AA\  were used respectively for \ce{MAPbI3} and \ce{FAPbI3} (Fig. S4, S5)\cite{mapi-structure,fapi-structure}. Relaxed unit cells were expanded into 2×2×2 supercells, each containing 96 atoms. The FA$_{0.8}$Cs$_{0.2}$PbI$_3$ supercell (89 atoms) was created by substituting a formamidinium (FA) group with a cesium (Cs) atom within an optimized 2×2×2 \ce{FAPbI3} supercell (originally 96 atoms) followed by variable cell relaxation (Fig. S6).
    
The slabs for \ce{MAPbI3}, \ce{FAPbI3}, and FA$_{0.8}$Cs$_{0.2}$PbI$_3$ were generated from relaxed bulk $2\times 2 \times 2$ supercells using pymatgen~\cite{Ong2013pymatgen} and atomic simulation environment (ASE)\cite{hjorthlarsenAtomicSimulationEnvironment2017} libraries. Structures were visualized using VESTA\cite{momma2011vesta} software. For each material, \ce{PbI2}-terminated symmetric slabs were used similar to previous works, simulating \ce{PbI2}-excess conditions\cite{doi:10.1021/jp511123s,FAPI-PbItermination,ahmadUnderstandingEffectLead2022,ahmadModulationPointDefect2024a}.  The thicknesses of the slabs were chosen to include 17 layers for \ce{MAPbI3} and \ce{FAPbI3}\cite{ahmadModulationPointDefect2024a} (Fig. S7, S8). To maintain symmetry of the slab, we used 19 layers for FA$_{0.8}$Cs$_{0.2}$PbI$_3$ (Fig. S9). A vacuum of 10~\AA\ was applied on both sides of the slab to remove interactions of periodic images. We used \hkl(001) surface slabs as in the previous work\cite{doi:10.1021/jp511123s,https://doi.org/10.1002/advs.202300056} as this surface has the lowest surface energy. In addition, we constructed a \hkl(100) surface terminated slab for FA$_{0.8}$Cs$_{0.2}$Pb(Br$_{x}$I$_{1-x}$)$_{3}$ with varying I/Br ratios.

All calculations employed the Perdew-Burke-Ernzerhof (PBE) functional\cite{perdewGeneralizedGradientApproximation1996} and norm-conserving pseudopotentials from the Quantum ESPRESSO Pseudo Dojo library.\cite{vansettenPseudoDojoTrainingGrading2018} Dispersion interactions were accounted for using Grimme's DFT-D3 correction method.\cite{grimmeConsistentAccurateInitio2010} The energy cutoff  for wave functions (ecutwfc) was consistently set to 80 Ry in all materials with k-point meshes of 3×3×3 for bulk calculations and 3×3×1 for slab calculations.
The force and energy convergence thresholds were set to $10^{-3}$ Ry/Bohr and $10^{-4}$ Ry. 

Defect formation energies (DFE) of Br antisite defects and I vacancy defects studied in this work were calculated using the standard formula\cite{freysoldtFirstprinciplesCalculationsPoint2014}:
\begin{equation}
\text{DFE} = E_{\mathrm{d}} - E_{\mathrm{p}} - \sum_i n_i \mu_i + q E_F + E_{\mathrm{corr}}
\end{equation}
where \( E_{\mathrm{d}} \) represents the calculated energy of the system containing the defect and \( E_{\mathrm{p}} \) denotes the total energy of the corresponding pristine system. The integer \( n_i \) indicates the number of atoms of species \( i \) either host or impurity that have been added (\( n_i > 0 \)) or removed (\( n_i < 0 \)) to form the defect. Each atomic species is associated with a chemical potential \( \mu_i \). The correction term was evaluated using sxdefectalign\cite{freysoldtFullyInitioFiniteSize2009} and sxdefectalign2D\cite{freysoldtFirstprinciplesCalculationsCharged2018} codes for bulk and slab respectively with dielectric constants obtained from density functional perturbation theory calculations. The total dielectric constants of MAPb(Br$_{0.5}$I$_{0.5}$)$_3$ and FA$_{0.8}$Cs$_{0.2}$Pb(Br$_{0.42}$I$_{0.58}$)$_3$ were found to be 30.55 and 50.53, respectively.

\section{Results and Discussion}

\subsection{Surface segregation tendency of halides}
To explain the origin and the mechanism of halide segregation in mixed halide perovskites, we analyzed slabs of the mixed halide, MAPb(Br$_{0.5}$I$_{0.5}$)$_3$ with three different distributions of Br and I near the surface and bulk layers: 1) uniform distribution of Br and I in all layers, 2) Br-enriched (I-depleted) surface layers, and 3) I-enriched (Br-depleted) surface layers. We performed DFT calculations using two settings: in the first (hereafter referred to as the constrained configuration), the middle layers are kept fixed to simulate bulk behavior, while in the second (referred to as the fully relaxed configuration), the middle layers are not fixed. The same DFT calculations were performed on another mixed halide, FA$_{0.8}$Cs$_{0.2}$Pb(Br$_{0.42}$I$_{0.58}$)$_{3}$ to compare its tendency for halide segregation. The relevant slab structures are shown in Fig. S11 and S12.

\begin{figure}[H]
    \centering
        \includegraphics[width=1\textwidth]{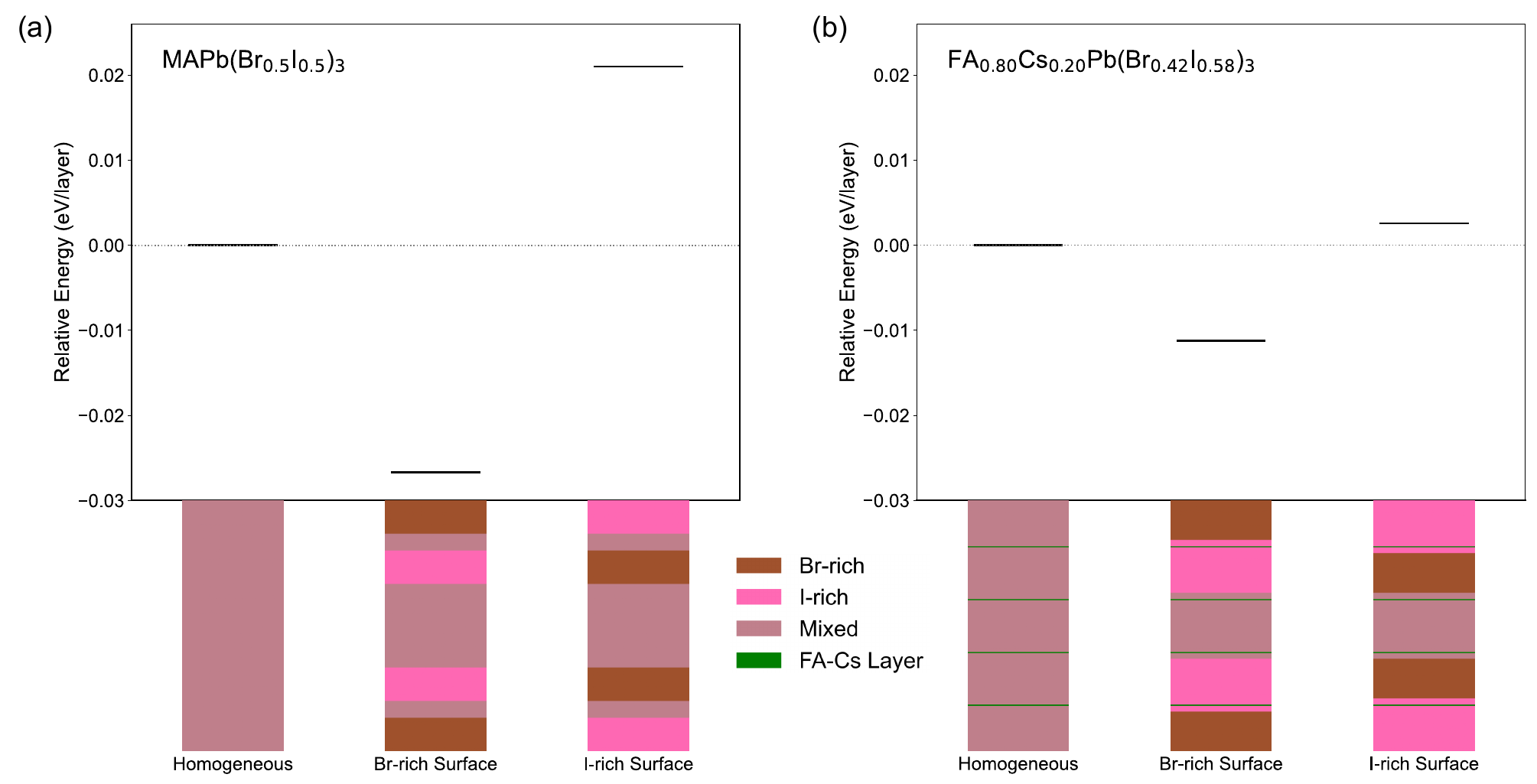}
    \caption{Energies of  (a) MAPb(Br$_{0.5}$I$_{0.5}$)$_3$ and (b) FA$_{0.8}$Cs$_{0.2}$Pb(Br$_{0.42}$I$_{0.58}$)$_3$ slabs in the constrained configuration where middle few layers were kept fixed. In both cases the homogeneous mixture is taken as reference state. The structure with a Br-enriched surface in both the materials (a and b) exhibits the lowest total energy, in contrast, surfaces enriched with I show the highest energies for both materials, indicating a thermodynamic preference for iodine to remain in the bulk rather than segregate to the surface. Homogeneous halide distributions fall in between.}
    \label{fig:energylevels}
\end{figure}

\autoref{fig:energylevels}a and b plot the energies of the three slabs for MAPb(Br$_{0.5}$I$_{0.5}$)$_3$ and FA$_{0.8}$Cs$_{0.2}$Pb(Br$_{0.42}$I$_{0.58}$)$_{3}$, respectively for the constrained configuration. The energies for the fully relaxed configurations are plotted in Fig. S1. We find that the slabs with Br-rich surfaces have the lowest energies in both constrained and fully relaxed configurations, while the slabs with the I-rich surface have the highest energies. The Br-rich surface slab is more stable compared to the homogeneous slab by 0.027 eV/layer for MAPb(Br$_{0.5}$I$_{0.5}$)$_3$ and 0.011 eV/layer for FA$_{0.8}$Cs$_{0.2}$Pb(Br$_{0.42}$I$_{0.58}$)$_{3}$. This result indicates that 1) halide segregation is thermodynamically preferred in mixed halide perovskites, leading to Br accumulation and I depletion near surfaces, and 2) the segregation tendency is lower in FA$_{0.8}$Cs$_{0.2}$Pb(Br$_{0.42}$I$_{0.58}$)$_{3}$ compared to MAPb(Br$_{0.5}$I$_{0.5}$)$_3$. The second observation explains the stability of FA$_{0.8}$Cs$_{0.2}$Pb(Br$_{0.42}$I$_{0.58}$)$_{3}$ to halide segregation compared to MAPb(Br$_{0.5}$I$_{0.5}$)$_3$, which agrees with the photoluminescence and X-ray diﬀraction measurements on FA$_{0.83}$Cs$_{0.17}$Pb(I$_{0.6}$Br$_{0.4}$)$_{3}$ by \citet{knightHalideSegregationMixedHalide2021}. While they explained the differences in halide segregation tendency due to A-site cation through the existence of low-barrier ionic pathways (kinetics), our calculations show that it is also thermodynamically more favorable for halide ions to segregate when the cation is MA compared to \ce{FA_{0.8}Cs_{0.2}}. Hence, A-site cation engineering can be a useful strategy for improving the thermodynamic stability of mixed halide perovskites\cite{knightHalideSegregationMixedHalide2021,rehman2017photovoltaic,zhou2018composition,braly2017current}.

Having established that Br tends to accumulate near the surfaces of mixed halide perovskites, we investigated the stability of Br antisite defects (Br$_\text{I}^{\times}$ in the Kr\"oger-Vink notation) with distance from the surface for three perovskite structures with different A-site cations, MA, FA and FA$_{0.8}$Cs$_{0.2}$ for the pristine halide composition of 0\% Br (pure I).
In each slab, one I atom was replaced by a Br atom to create the antisite defect in different layers starting from the surface layer to the middle layer (Fig. S7, Fig. S8, Fig. S9). The middle layer in each slab should exhibit similar characteristics as the bulk.

\begin{figure}[H]
    \centering
        \includegraphics[width=\textwidth]{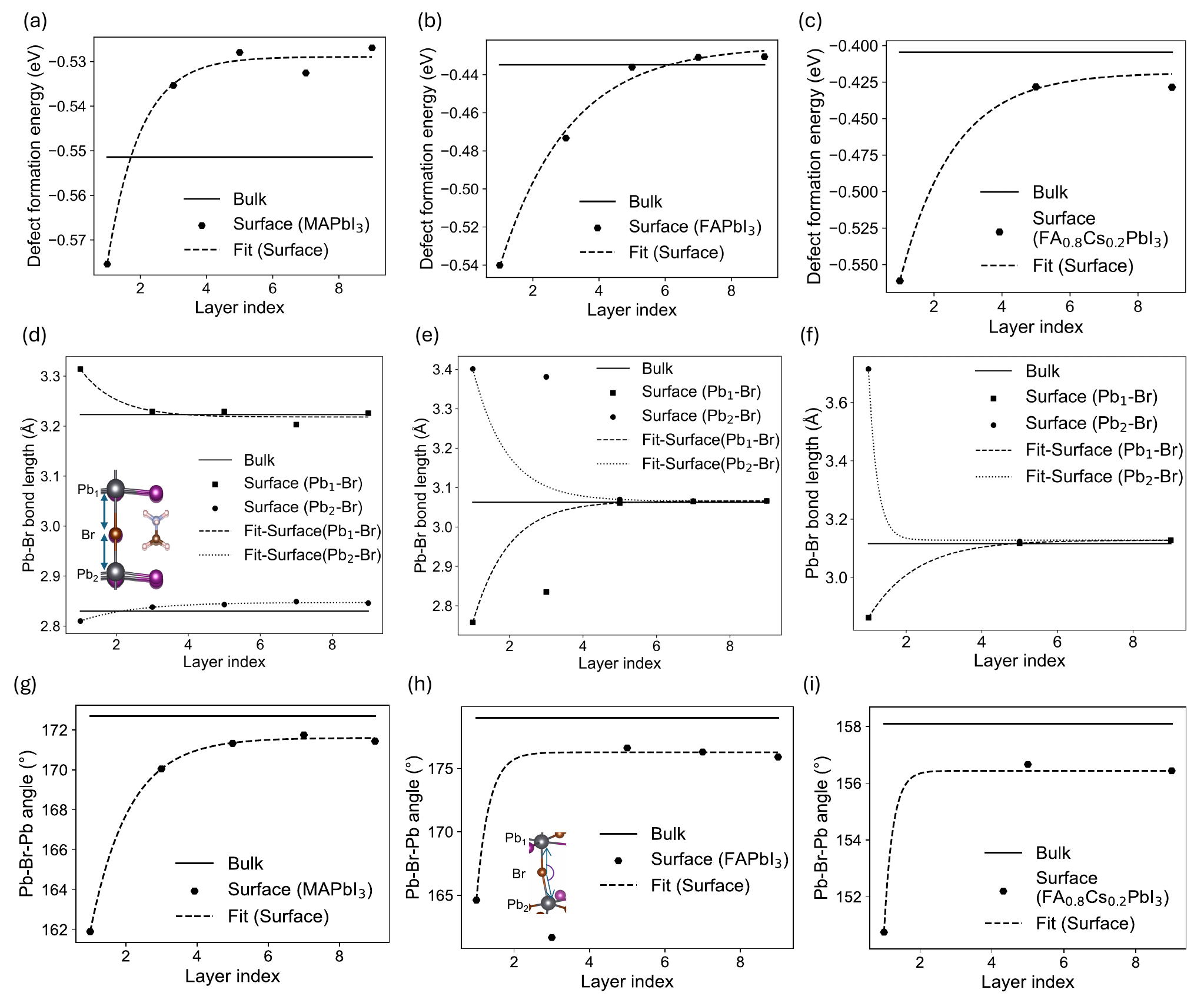}
    \caption{Depth-dependent properties of the Br antisite defect. For all panels, data are plotted against the layer index (0 = surface). (a-c) Variation in DFE as a function of depth from the surface is presented for (a) 17-layer \ce{MAPbI3}, (b) 17-layer \ce{FAPbI3}, and (c) 19-layer FA$_{0.8}$Cs$_{0.2}$PbI$_3$ slabs. In all cases, the DFE is substantially lower near the surface compared to the bulk, indicating a thermodynamic preference for Br to localize at the surface. An exponential increase in DFE with depth is observed. (d-f) Pb–Br bond lengths in the presence of a Br antisite defect for (d) \ce{MAPbI3},(e) \ce{FAPbI3} and (f) FA$_{0.8}$Cs$_{0.2}$PbI$_3$. The two different bond lengths are shown in the inset of (d). Different Pb–Br-Pb bond angles in the presence of a Br antisite defect for (g) \ce{MAPbI3},(h) \ce{FAPbI3} and (i) FA$_{0.8}$Cs$_{0.2}$PbI$_3$.  The different angles are shown in the inset of (h). The variation in bond lengths across layers reflects how structural relaxation influences the DFE, causing it to converge toward the bulk value. }
    \label{fig:dfe}
\end{figure}

\autoref{fig:dfe}a, b, and c plot the variation in DFE with distance from the surface for 0\% Br (pure I) composition for  \ce{MAPbI3}, \ce{FAPbI3}, and FA$_{0.8}$Cs$_{0.2}$PbI$_3$, respectively. In each case, we find that the antisite DFE is negative for all the layers, indicating a thermodynamic preference for replacement of I atoms by Br when the perovskite is in equilibrium with \ce{I2} and \ce{Br2} gases. This is consistent with the loss of I atoms (as opposed to Br atoms) as \ce{I2} gas in mixed halide perovskites containing Br and I during illumination\cite{leeEvidence2Loss2025}. In the bulk, the DFE follows the trend: MA $<$ FA $<$ FA$_{0.8}$Cs$_{0.2}$, indicating strongest driving force for I replacement by Br in \ce{MAPbI3}. 
The DFE is minimum when the antisite defect is in the surface layer and increases with distance from the surface, following an exponential behavior. Eventually, the DFE converges to the bulk value with a small discrepancy of $< 0.02$ eV between the converged slab DFE and the bulk DFE.

To explain the behavior of the DFE near the surface, we examined the local bonding environment and distortion around the defect after ionic relaxation. We calculated the Pb-Br bond lengths and the Pb-Br-Pb bond angles with distance from the surface.
\autoref{fig:dfe}(d-f) shows the Pb-Br bond lengths associated with the two nearest neighbor Pb atoms coordinated to the Br antisite defect  as shown in the inset of ~\autoref{fig:dfe}d. The difference between the two Pb-Br bond lengths increases on moving closer to the surface. 
For all three cases the Pb-Br bond lengths exhibit the same trend as the DFE. The progressively asymmetric behaviour of Pb–Br bond lengths near the surface makes it easier for defects to form. (Fig. S3). The bond length asymmetry in the slab also explains the difference between the converged slab DFE and the bulk DFE in \ce{MAPbI3}. This near-surface distortion is also reflected in the Pb-Br-Pb angle shown in the inset of ~\autoref{fig:dfe}h, which shows a larger deviation from the bulk value (enhanced octahedral tilting) near the surface, decreasing towards the bulk (\autoref{fig:dfe}(g-i)). The correlation between the Pb-Br-Pb bond angles and perovskite properties has been observed in other studies as well~\cite{janaStructuralDescriptorEnhanced2021,luo2025synergistic}. 
Beyond the nearest neighbor to Br antisite defect (Pb), the I-Br distances also follows a similar trend as the DFE (Fig.~S2).

We fitted the DFE variation to the exponential form similar to our previous work~\cite{ahmadModulationPointDefect2024a,limonHeterogeneityPointDefect2024c}, $\text{DFE}(x) = A + B \exp(-x/l)$ where $x$ is the distance of defect from the surface and $A$, $B$, and $l$ are constants~\cite{ahmadModulationPointDefect2024a}. Here, $l$ is the decay length which is a quantitative measure of how far the influence of a surface extends into the bulk of the perovskite.  For \ce{MAPbI3}, the decay length was found to be approximately 4.2 {\AA}, indicating a very sharp transition from surface to bulk behavior. In contrast, \ce{FAPbI3} and FA$_{0.8}$Cs$_{0.2}$PbI$_3$ showed more gradual transitions with decay lengths of 8.9 {\AA} and 6.6 {\AA}, respectively. These values suggest that in \ce{MAPbI3}, the influence of surface on defect energetics is highly localized to the first layer or two, while in \ce{FAPbI3}, the surface effects persist deeper into the slab. 

\begin{figure}[H]
    \centering
        \includegraphics[width=1\textwidth]{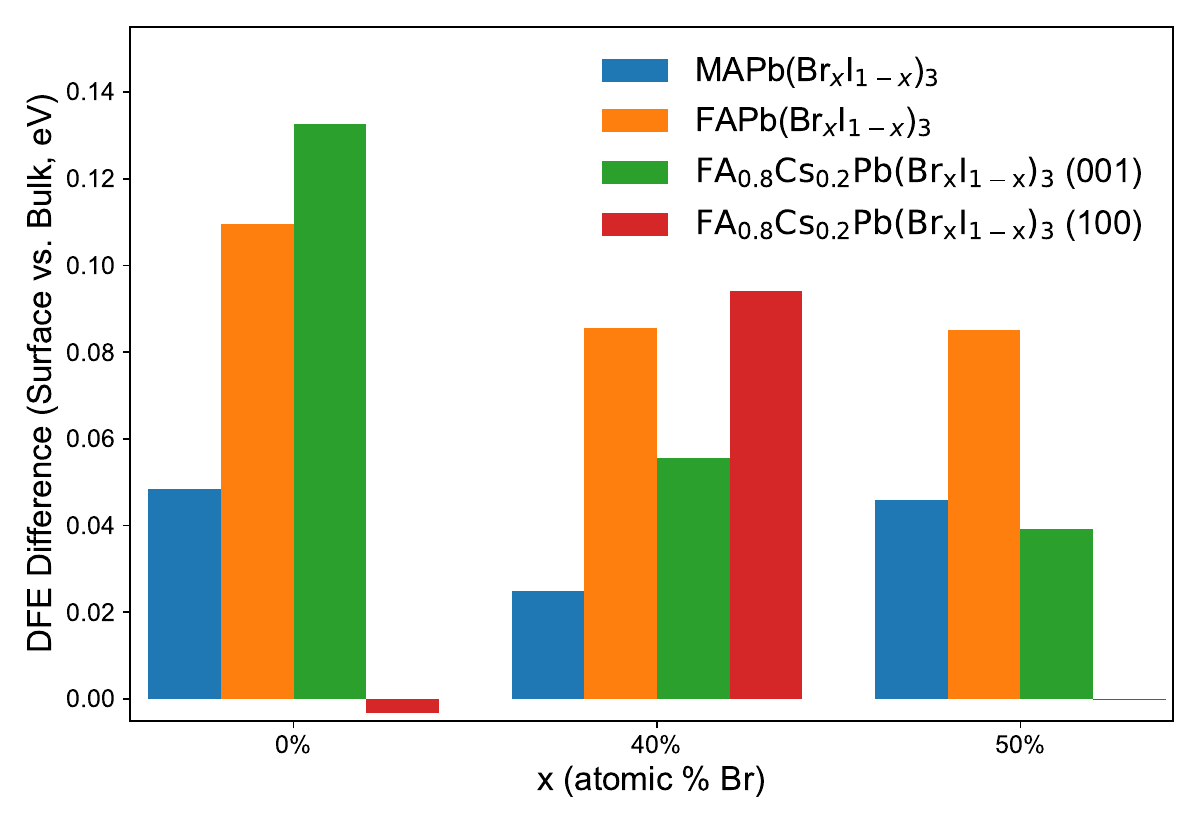}
    \caption{Change in DFE from bulk to surface (DFE$_{bulk}$-DFE$_{surface}$) for various perovskite compositions as a function of Br concentration ($x$). The DFE difference depends on the Br/I ratio in each structure. At 0\% Br, MAPbI$_3$ shows the highest resistance to surface segregation. However, as Br content increases, FA$_{0.8}$Cs$_{0.2}$Pb(Br$_x$I$_{1-x}$)$_3$ exhibits progressively greater resistance, reaching the highest stability at a 1:1 Br:I ratio.}
    \label{fig:dfecomp}
\end{figure}

The key thermodynamic descriptor for halide segregation towards surfaces is the difference in the DFE of Br$_\text{I}$ between the bulk and the surface, $\Delta \text{DFE} = \text{DFE}_{\text{bulk}} - \text{DFE}_{\text{surface}}$. The greater the difference, the higher the tendency for Br to segregate near surfaces. We calculated $\Delta$DFE for the slab for three different pristine halide compositions, 0\% Br (pure I), 40\% Br(Fig. S10) and 50\% Br where Br atoms were uniformly distributed within the slab by evenly substituting I. 
\autoref{fig:dfecomp} compares the values of $\Delta$DFE for the three compositions. $\Delta$DFE is highest for pure I structures with FA$_{0.8}$Cs$_{0.2}$PbI$_3$ exhibiting the largest value (0.133 eV). This corresponds to a surface defect density enhancement ($c_{surface}/c_{bulk}$) of $\sim$171.515 times  compared to the bulk at 300 K, based on the relation $c \propto e^{-\Delta \mathrm{DFE} / (k_B T)}$, where $c$ is the defect density, $k_B$ is the Boltzmann constant, and $T$ is the temperature. In contrast, \ce{MAPbI3} shows a more moderate enhancement of $\sim$6.5 times under the same conditions. 
For the 40\% Br composition, FAPb(Br$_{0.4}$I$_{0.6}$)$_3$ shows the highest surface segregation tendency, while MAPb(I$_{0.6}$Br$_{0.4}$)$_3$ still exhibits highest resistance to segregation. Notably, at equimolar 50\% Br composition, FA$_{0.8}$Cs$_{0.2}$Pb(I$_{0.5}$Br$_{0.5}$)$_3$ shows the lowest $\Delta$DFE of 0.04 eV, indicating the highest resistance to surface segregation in agreement with experiments\cite{knightHalideSegregationMixedHalide2021}. At this composition, the calculated enhancement of defect density is only 4.5 for FA$_{0.8}$Cs$_{0.2}$Pb(I$_{0.5}$Br$_{0.5}$)$_3$, compared to 10.03 for MAPb(Br$_{0.5}$I$_{0.5}$)$_3$. 
In addition, we evaluated the segregation tendency of \hkl(100)-oriented FA$_{0.8}$Cs$_{0.2}$Pb(I$_{0.5}$Br$_{0.5}$)$_3$ slab, which exhibits a higher surface energy (0.27 J/m$^2$) compared to the \hkl(001) surface (0.096 J/m$^2$) to study the effect of crystallographic orientation. We obtained nearly zero $\Delta$DFE for this slab, suggesting no thermodynamic driving force for surface segregation although this slab showed lower resistance for the 40\% Br composition case.
These results reveal that the A-site cation, halide composition, and the surface orientation critically influence the defect thermodynamics and halide segregation in mixed halide perovskites. The suppression of halide segregation with FA$_{0.8}$Cs$_{0.2}$ as the A-site cation for the equimolar composition positions it as a favorable candidate for stable mixed halide perovskite devices.

\subsection{Halide segregation under illumination}
While defect thermodynamics provides the driving force for segregation in mixed halide perovskites, the process requires rapid ion rearrangement facilitated by high ion conductivity. The process is kinetically initiated by light which is known to enhance the ion conductivity of perovskites by several orders of magnitude~\cite{kimLargeTunablePhotoeffect2018}. In the dark, the perovskite reverts to the mixed halide state due to entropic forces~\cite{Dubose-kamat-holetrapping}, which are of the same order as the thermodynamic descriptor for segregation, $\Delta $DFE ($\sim k_BT$). Light generates holes which cause oxidation of the halide species, resulting in the production of neutral halide atoms~\cite{KERNER20212273,photolysis}. The oxidation will happen to a different degree for \ce{I-} and \ce{Br-}~\cite{kimLargeTunablePhotoeffect2018,senocrateSolidStateIonicsHybrid2019}. The neutral halide atoms  are readily displaced from their lattice positions to the interstitial ones to share the charge of halide ions, creating halide vacancies in the process. Furthermore, the halide atoms can be lost as \ce{I2} or \ce{Br2} gas through the perovskite surface under sustained illumination, which cannot be reversed in the dark.
Overall, these mechanisms result in segregation of halide ions under illumination. Two important properties that determine rates of segregation are the 1) tendency of holes to oxidize halide ions and 2) tendency to form halide vacancies. The differences in these properties for \ce{Br-} and \ce{I-} lead to halide segregation.  Next, we quantify the former property by the localization of holes near specific halide ions and the latter by the DFE of halide vacancies.

\begin{figure}[H]
    \centering

        \centering
        \includegraphics[width=1\textwidth]{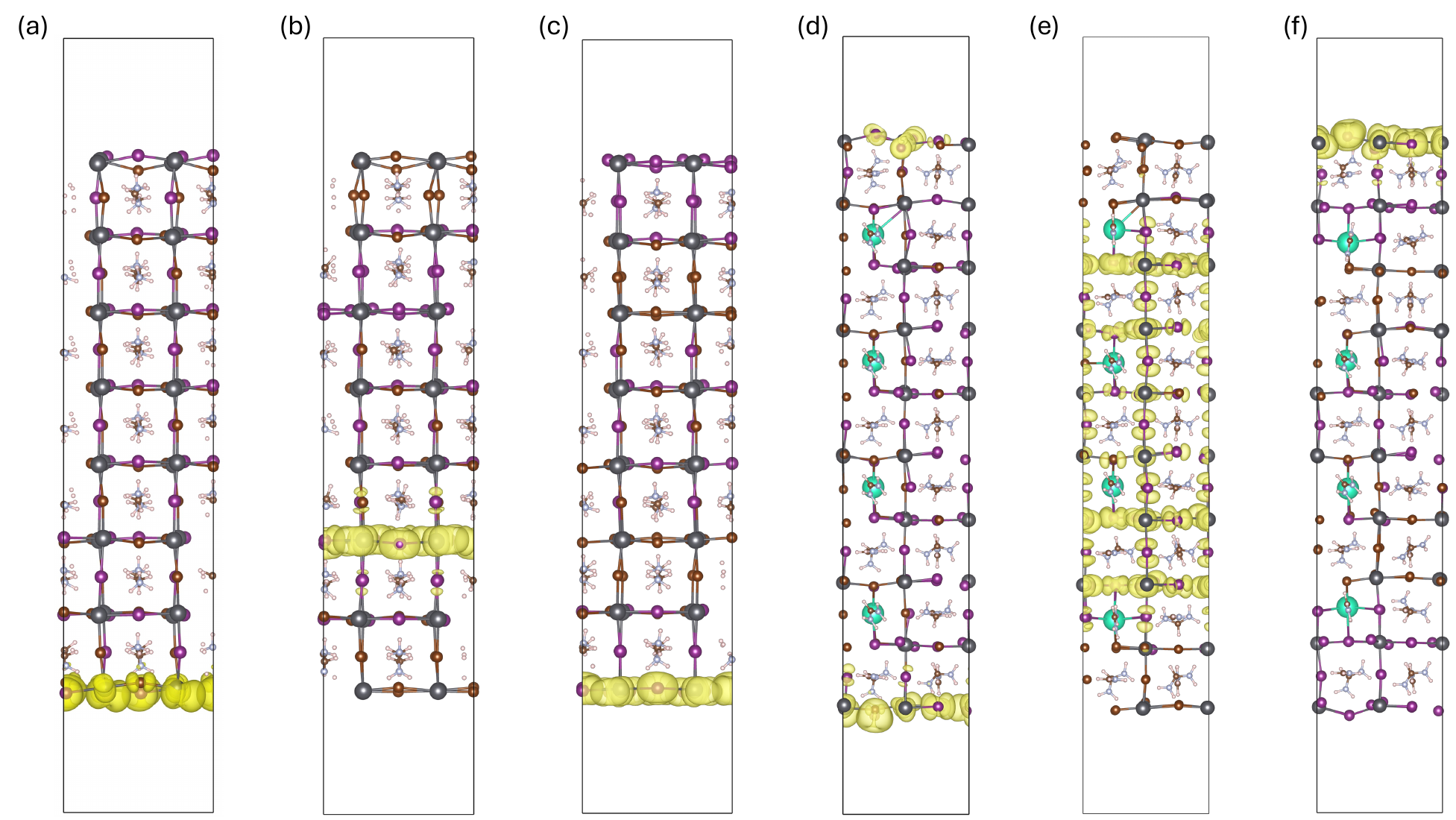}

    \caption{Position of VBM in different materials. (a-c) VBM position in different types of slabs of MAPb(Br$_{0.5}$I$_{0.5}$)$_3$. (a)Homogeneous mixture of I and Br. (b)Br atoms are near the surfaces.(c)I atoms are near the surfaces. (d-f)VBM position in different types of slabs of FA$_{0.8}$Cs$_{0.2}$Pb(Br$_{0.42}$I$_{0.58}$)$_3$ . (d) Homogeneous mixture of I and Br. (e) Br atoms are near the surfaces. (f) I atoms are near the surfaces. In all surface-segregated models (b, c, e, f), the middle 5 layers are kept fixed.}
    \label{fig:vbmcbm}
\end{figure}

\noindent \textbf{Hole localization in mixed halide perovskites}. 
Due to the importance of holes in facilitating halide segregation, we performed DFT calculations to determine hole positions within MAPb(Br$_{0.5}$I$_{0.5}$)$_3$ and FA$_{0.8}$Cs$_{0.2}$Pb(Br$_{0.42}$I$_{0.58}$)$_3$ slabs. 
~\autoref{fig:vbmcbm} illustrates the results of hole localization simulations within these slabs. Panels (a–c) represent MAPb(Br$_{0.5}$I$_{0.5}$)$_3$ slabs with (a) uniform distribution of Br and I, (b) Br-rich surface, and (c) I-rich surface. The yellow isosurfaces represent the valence band maximum (VBM), indicating regions of hole accumulation. In (a), the holes accumulate near the surface, while in (b) and (c), the holes accumulate in the I-rich regions, i.e., within the bulk for Br-rich surface slab and at the surface in I-rich surface slab. The hole positions for the FA$_{0.8}$Cs$_{0.2}$Pb(Br$_{0.42}$I$_{0.58}$)$_3$ slab are similar but more distributed as shown in ~\autoref{fig:vbmcbm}(d-f). Our results agree with other studies\cite{Dubose-kamat-holetrapping, hoffman2016transformation,dubose2021modulation}, which show that holes preferentially accumulate in iodide-rich regions due to significant valence band offsets between domains containing I and Br.\cite{kamat-TiO2} This preferential localization of holes acts as a critical driving force for phase segregation, promoting further ion migration and contributing to the formation of recombination centers in iodide-rich domains.

\noindent \textbf{Halide vacancy formation in mixed halide perovskites}. Halide vacancies are known to enhance ion migration, which drives halide segregation. Halide-deficient perovskite films exhibit higher halide vacancy concentration, which increases halide segregation~\cite{yoonShiftHappensHow2017}. Furthermore, facile vacancy formation can trigger the irreversible loss of halogens from the perovskite surface.

\begin{figure}[H]
    \centering

        \centering
        \includegraphics[width=1\textwidth]{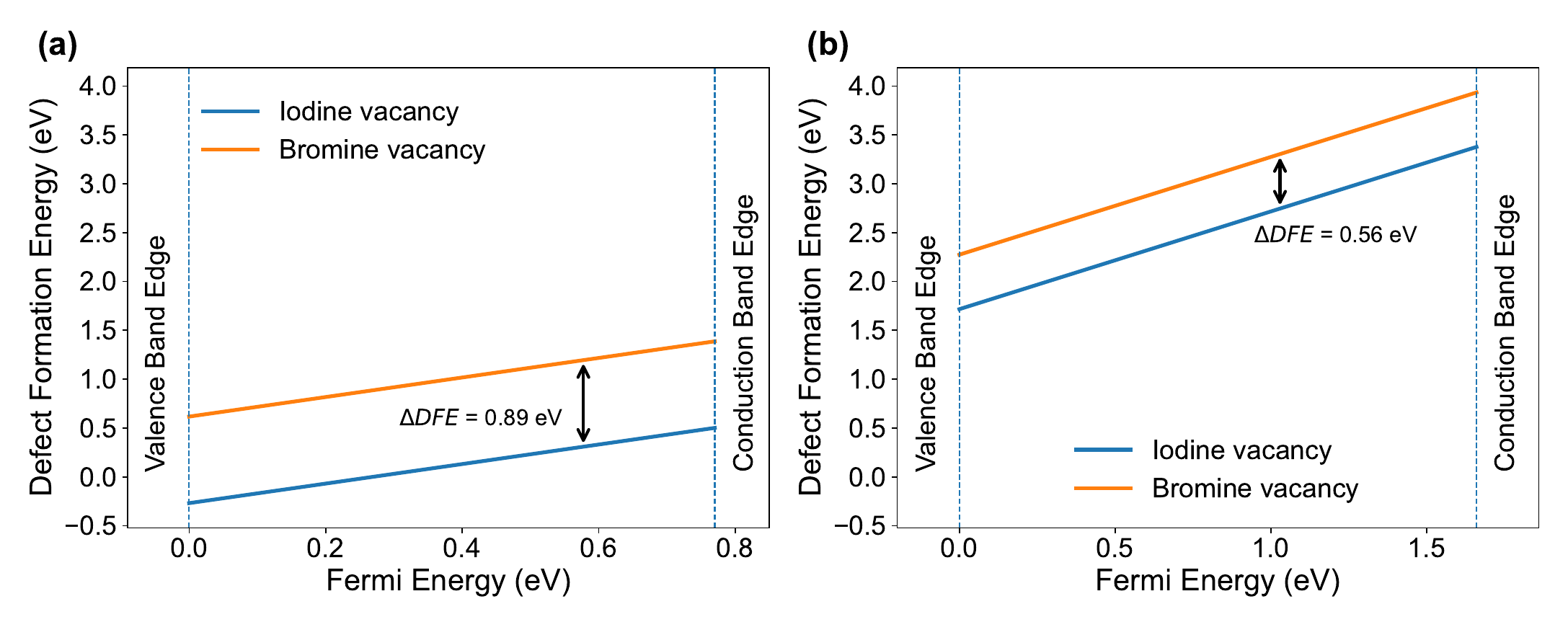}
    
    \caption{Defect formation energies of I ($V_\mathrm{I}^{\bullet}$) and Br ($V_\mathrm{Br}^{\bullet}$) vacancies at the surface of (a) MAPb(Br$_{0.5}$I$_{0.5}$)$_3$ and (b) FA$_{0.8}$Cs$_{0.2}$Pb(Br$_{0.50}$I$_{0.50}$)$_3$ as a function of Fermi level across the bandgap. The valence band maximum (VBM) and conduction band minimum (CBM) are indicated by dashed lines. In both materials, $V_\mathrm{I}^{\bullet}$ exhibits consistently lower formation energy than $V_\mathrm{Br}^{\bullet}$ over the full Fermi level range, indicating a stronger thermodynamic preference for $V_\mathrm{I}^{\bullet}$ formation at the surface.}

    \label{fig:vacancy}
\end{figure}

To evaluate the differences in halide vacancy formation, we constructed slabs of MAPb(Br$_{0.5}$I$_{0.5}$)$_3$ and FA$_{0.8}$Cs$_{0.2}$Pb(I$_{0.50}$Br$_{0.50}$)$_{3}$ containing either an I vacancy ($V_{\mathrm{I}}^{\bullet}$) or a Br vacancy ($V_{\mathrm{Br}}^{\bullet}$). We then evaluated their DFEs across the full Fermi-level range. As shown in \autoref{fig:vacancy}, For both perovskites, $V_{\mathrm{I}}^{\bullet}$ consistently exhibits a lower DFE than $V_{\mathrm{Br}}^{\bullet}$ in both perovskites, indicating that the formation of I vacancy is more energetically favorable. The difference in DFE is 0.89 eV for MAPb(Br$_{0.5}$I$_{0.5}$)$_3$ and 0.56 eV for FA$_{0.8}$Cs$_{0.2}$Pb(I$_{0.50}$Br$_{0.50}$)$_{3}$. This trend persists in the bulk albeit with smaller formation energy differences (Fig. S13). The observed trend in DFEs is consistent with experimental observations showing that mixed I-Br perovskites preferentially lose I as volatile $\mathrm{I}_2$ at the perovskite-air interface under illumination, while no similar Br loss has been reported.\cite{leeEvidence2Loss2025} Moreover, the absolute DFEs in FA$_{0.8}$Cs$_{0.2}$Pb(I$_{0.50}$Br$_{0.50}$)$_{3}$ are higher than those in MAPb(Br$_{0.5}$I$_{0.5}$)$_3$, suggesting that vacancy formation is generally less favorable in FA$_{0.8}$Cs$_{0.2}$Pb(I$_{0.50}$Br$_{0.50}$)$_{3}$. This finding supports the notion that FA$_{0.8}$Cs$_{0.2}$Pb(I$_{0.50}$Br$_{0.50}$)$_{3}$  exhibits enhanced resistance to halide-loss induced instability compared to MAPb(Br$_{0.5}$I$_{0.5}$)$_3$.

\begin{figure}[htbp]
    \centering
        \includegraphics[width=\textwidth]{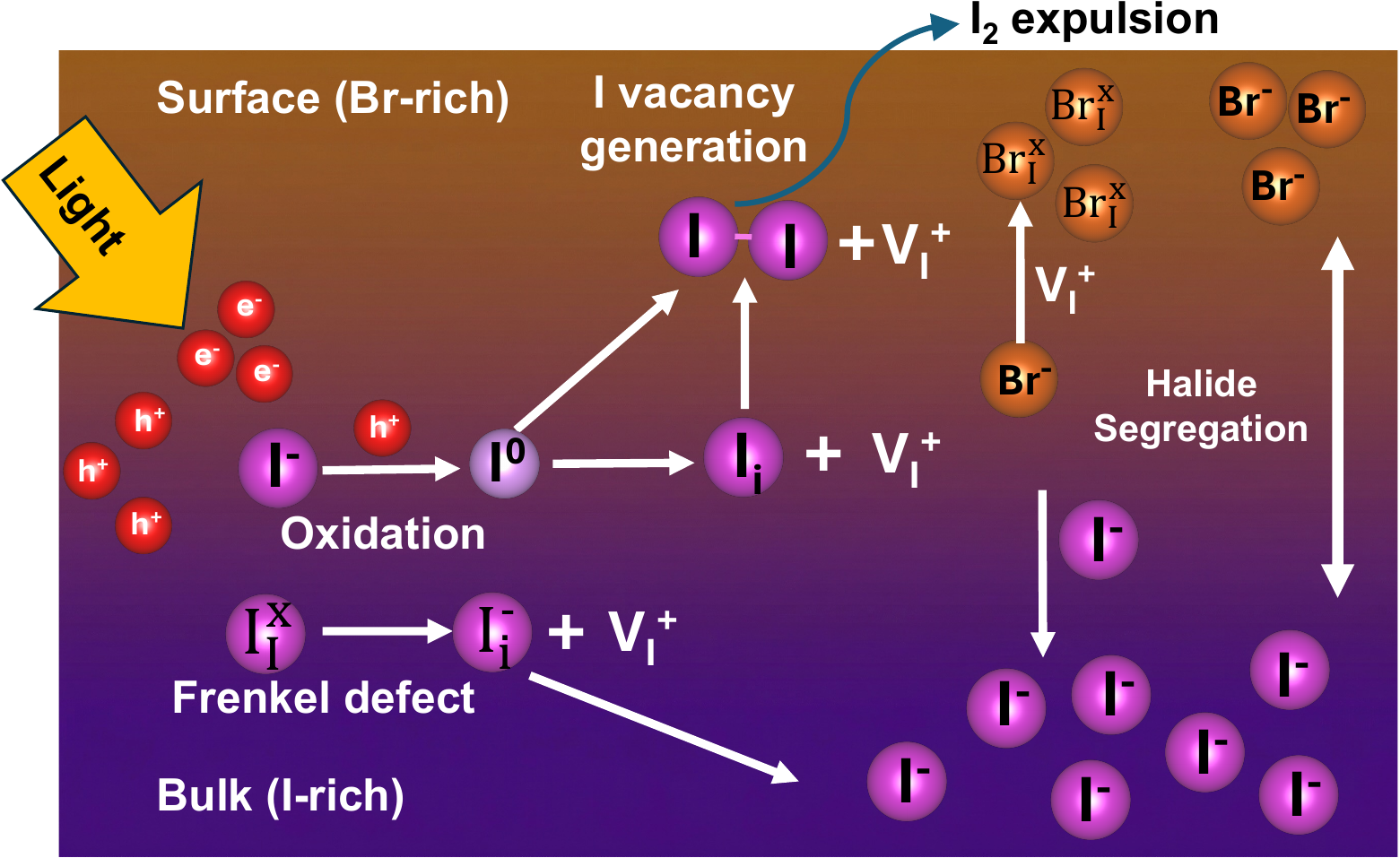}
    
    \caption{Proposed mechanism of halide segregation under illumination. Illumination opens up pathways for generation of iodide vacancies which enhance ion conductivity of the perovskite, enabling structural rearrangements required for movement of Br towards the surface and I towards the bulk.}

    \label{fig:mechanism}
\end{figure}

\noindent \textbf{Mechanism}. Based on the calculated properties, we propose a mechanism for light-induced halide segregation  as illustrated in \autoref{fig:mechanism}. Light activates several pathways for generation of iodine vacancies that enhance ion conductivity of the perovskite for structural rearrangement: 1) selective iodide oxidation that produces neutral iodine species, which are subsequently expelled as \ce{I2}, 2) displacement of neutral iodine from lattice site to interstitial site, 3) formation of iodine Frenkel defects at the surface due to low formation energies~\cite{meggiolaroFormationSurfaceDefects2019,ahmadUnderstandingEffectLead2022}. The generated highly mobile iodine vacancies migrate toward the surface while iodide ions move toward the bulk. Br ions then occupy these vacancies to form Br antisite defects, resulting in Br enrichment and I depletion at the surface. We note that I may also be expelled as \ce{I-} in solution~\cite{doi:10.1021/jacs.9b04568}.

\section{Conclusions}
In conclusion, we report the intrinsic structural and thermodynamic origins of halide segregation in mixed halide perovskites, mediated by halide antisite and vacancy defects. Our results demonstrate that \ce{Br} exhibits a thermodynamic preference for surface sites over \ce{I}, providing a fundamental driving force for segregation. This driving force is modulated by the A-site cation, with FA$_{0.8}$Cs$_{0.2}$ exhibiting the smallest value at 50\% Br molar concentration due to small energy difference between surface and bulk Br. The tendency for segregation is also influenced by the halide ratio within the perovskite. We connect the defect thermodynamics to light-driven kinetics by showing that photogenerated holes localize at iodide sites, thereby selectively oxidizing them. This process accelerates halide vacancy generation, promoting segregation and potential irreversible degradation via \ce{I2} loss.

Our work establishes structural and compositional descriptors for designing perovskites that are resistant to halide segregation. While A-site cations are known to influence stability of mixed halide perovskites, our work clarifies their role through the proposed descriptor: the difference in DFE of Br antisite defects between the surface and the bulk. This descriptor is governed by the local bonding environment around Br, specifically the Pb-Br bond lengths and the Pb-Br-Pb bond angles, describing the degree of octahedral distortion. To minimize halide segregation, the perovskite lattice and its interfaces must be engineered to prevent the energy asymmetry of halide ions between the bulk and interfaces. While this work focuses on the perovskite-air interface, future studies should quantify these effects at other critical interfaces, including grain boundaries and junctions with charge transport layers.

\begin{acknowledgement}
We thank M. S. R. Limon for assistance with charged defect correction calculations. We acknowledge alternative energy research funding from Edward E. Whitacre Jr. College of Engineering and Mechanical Engineering Department startup grant at Texas Tech University for supporting this research. 
We also acknowledge the High-Performance Computing Center (HPCC) at Texas Tech University and the Lonestar6 research allocation (DMR24029 and DMR25016) at the Texas Advanced Computing Center (TACC) for providing computational resources that have contributed to the research results reported in this paper.

\end{acknowledgement}
\begin{suppinfo}

Details of computational methods, structural information and additional analyses conducted.

\end{suppinfo}

\bibliography{zotero,refs,refs_nav}



\section*{Supplementary material}

# Supporting Information: Understanding halide segregation in metal halide perovskites through defect thermodynamics

Abrar Fahim Navid and Zeeshan Ahmad*

*Department of Mechanical Engineering, Texas Tech University, Lubbock, Texas 79409, USA*

E-mail: zeeahmad@ttu.edu

## Calculation of Surface Energy:

To calculate the surface energy of the PbI$_2$ terminated symmetric slab of FA$_{0.8}$Cs$_{0.2}$Pb(Br$_{0.5}$I$_{0.5}$)$_3$ (001) and FA$_{0.8}$Cs$_{0.2}$Pb(Br$_{0.5}$I$_{0.5}$)$_3$ (100), we used the following equation:[1]

$$\text{surface energy} = \frac{1}{2A}\left(E_{slab} - n_{\text{unit}} \cdot E_{\text{unit}} + \sum_i n_i \cdot \mu_i\right) \qquad (1)$$

Here, A is the surface area facing the vacuum of one side of the slab. E$_{\text{slab}}$ is the total energy of each symmetric slab, n$_{\text{unit}}$ is the integer number of FACsPb(Br$_{0.5}$I$_{0.5}$)$_3$ unit cells present in the slab, E$_{\text{unit}}$ is the total energy of one unit cell, n$_i$ is the number of atoms of type i present in the slab in excess of the stoichiometric amount based on the reference bulk composition and µ$_i$ is the chemical potential of element i.

The chemical potential of atoms FA,Cs ,I and Br were found using the process stated in previous literatures.[2,3]

# Lattice Parameters : For MAPbI$_3$ ,FAPbI$_3$ and FA$_{0.8}$Cs$_{0.2}$PbI$_3$ unit cells

Table S1: Lattice Parameters for MAPbI$_3$ ,FAPbI$_3$ and FA$_{0.8}$Cs$_{0.2}$PbI$_3$ [4,5]

| Lattice Parameters | FAPbI$_3$ (Å) | MAPbI$_3$ (Å) | FA$_{0.8}$Cs$_{0.2}$PbI$_3$ (Å) |
|---|---|---|---|
| $a$ | 6.276 | 6.362 | 12.728 |

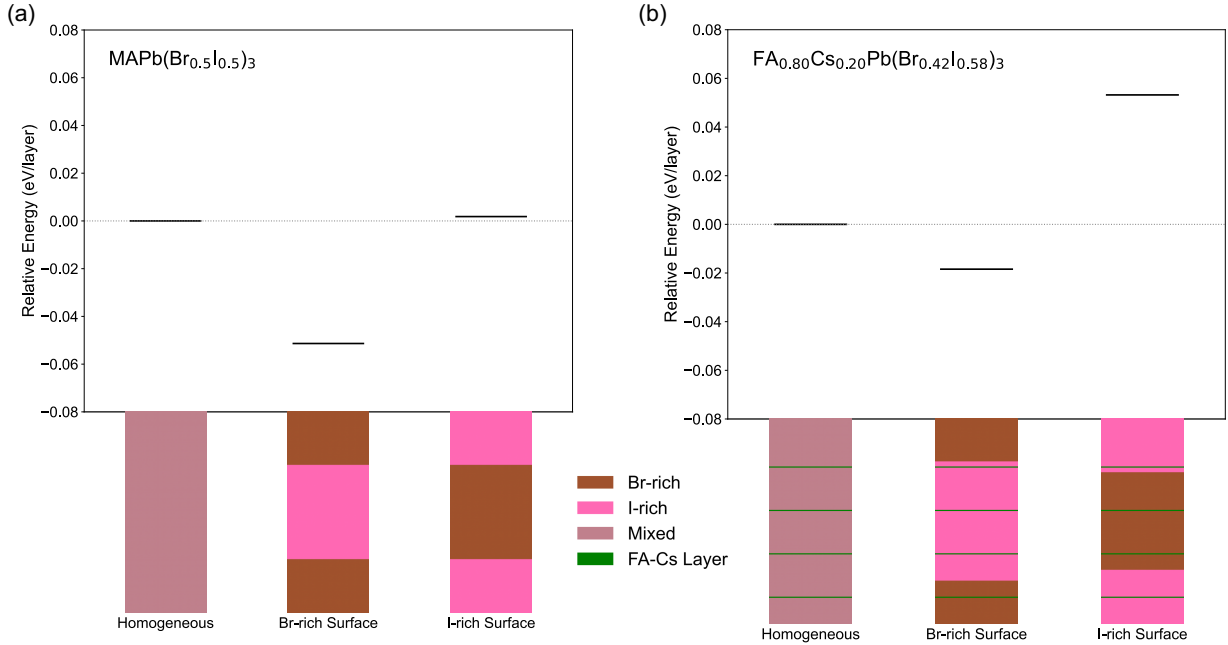

Figure S1: Energy profiles of various (a)MAPb(Br$_{0.5}$I$_{0.5}$)$_3$ and (b) FA$_{0.8}$Cs$_{0.2}$Pb(Br$_{0.42}$I$_{0.58}$)$_3$ slab configurations with no fixed middle layers. In both cases the homogeneous mixture is taken as reference state. The structure with a Br-enriched surface in both the materials (a and b) exhibits the lowest total energy, in contrast, surfaces enriched with iodine show the highest energies for both materials, indicating a thermodynamic preference for iodine to remain in the bulk rather than segregate to the surface. Homogeneous halide distributions fall in between.

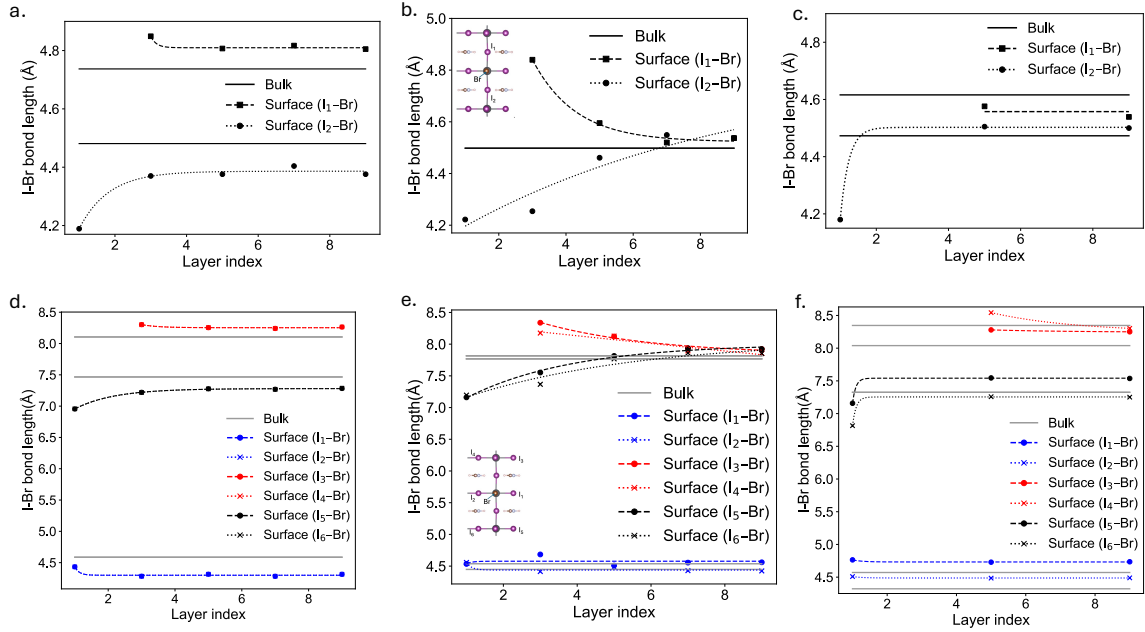

Figure S2: Different I–Br bond lengths in the presence of a Br antisite defect are plotted against the layer index (0 = surface) for (a,d) MAPbI$_3$, (b,e) FAPbI$_3$ and (c,f) FA$_{0.8}$Cs$_{0.2}$PbI$_3$ (a-c) Shows the bond distances for the Br to the nearest neighbor I and (d-f) shows the bond distances for the Br to the next nearest neighbor I. The different bond lengths are shown in the inset of b and d.

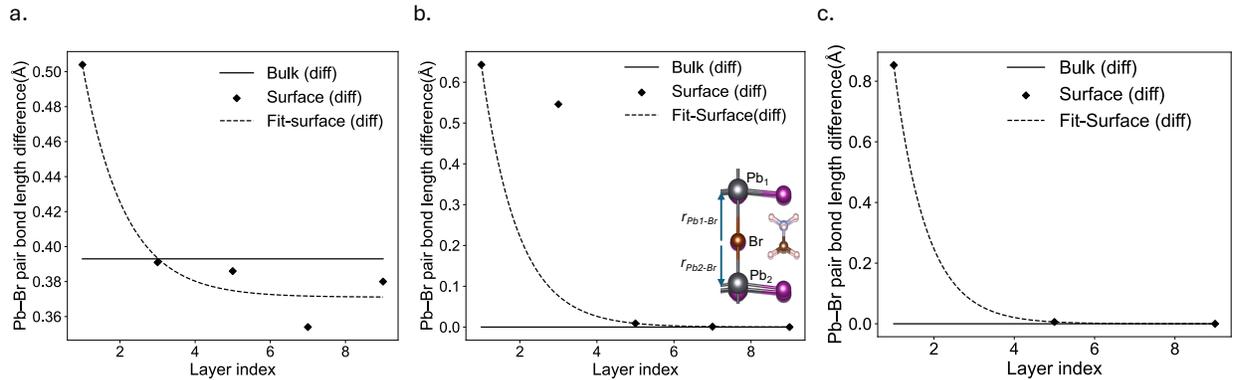

Figure S3: Pb-Br bond-length asymmetry, defined as $\Delta r_{Pb-Br} = |r_{Pb_1-Br} - r_{Pb_2-Br}|$, as a function of layer index (layer 0 corresponds to the surface) for a Br antisite defect in (a) MAPbI$_3$, (b) FAPbI$_3$, and (c) FA$_{0.8}$Cs$_{0.2}$PbI$_3$. An inset in (b) illustrates the Pb$_1$-Br and Pb$_2$-Br bonds used to compute $\Delta r_{\text{Pb-Br}}$.

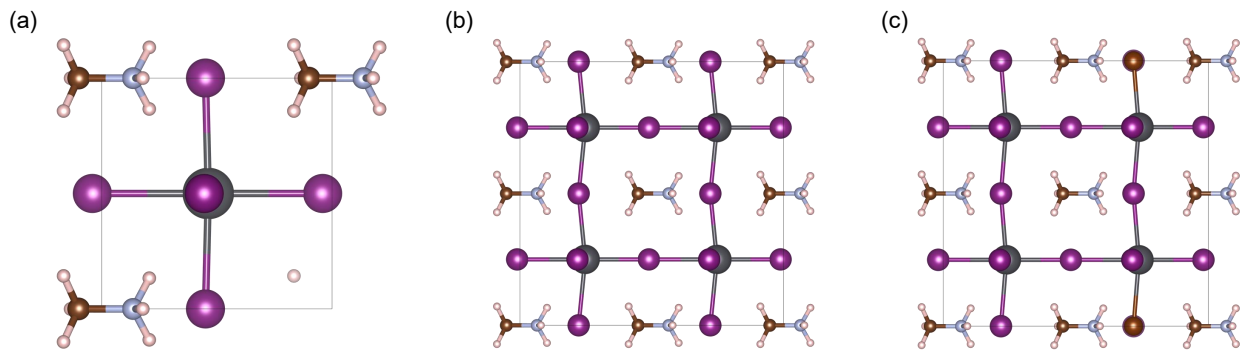

Figure S4: Structures of MAPbI$_3$ (a)Unit cell ,(b) 2 by 2 by 2 supercell and (c) One Br anti-site defect in a 2 by 2 by 2 supercell.

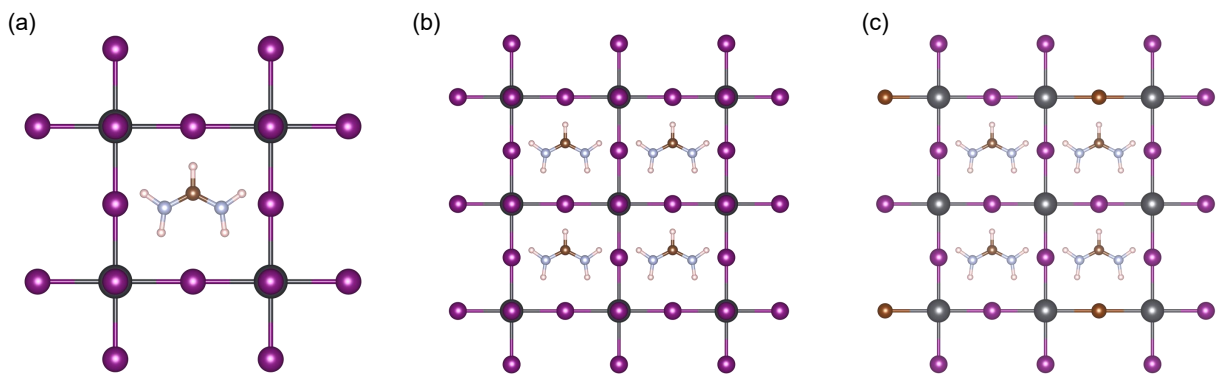

Figure S5: Structures of FAPbI$_3$ (a)Unit cell ,(b) 2 by 2 by 2 supercell and (c) One Br anti-site defect in a 2 by 2 by 2 supercell.

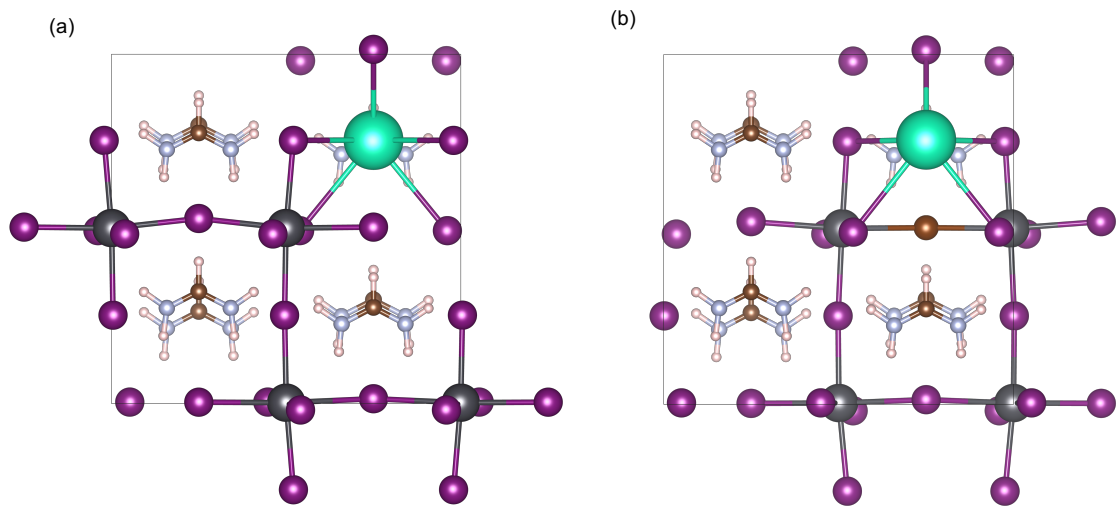

Figure S6: Structures of $FA_{0.8}Cs_{0.2}PbI_3$ (a) Unit cell and (b) One Br anti-site defect in the unit cell.

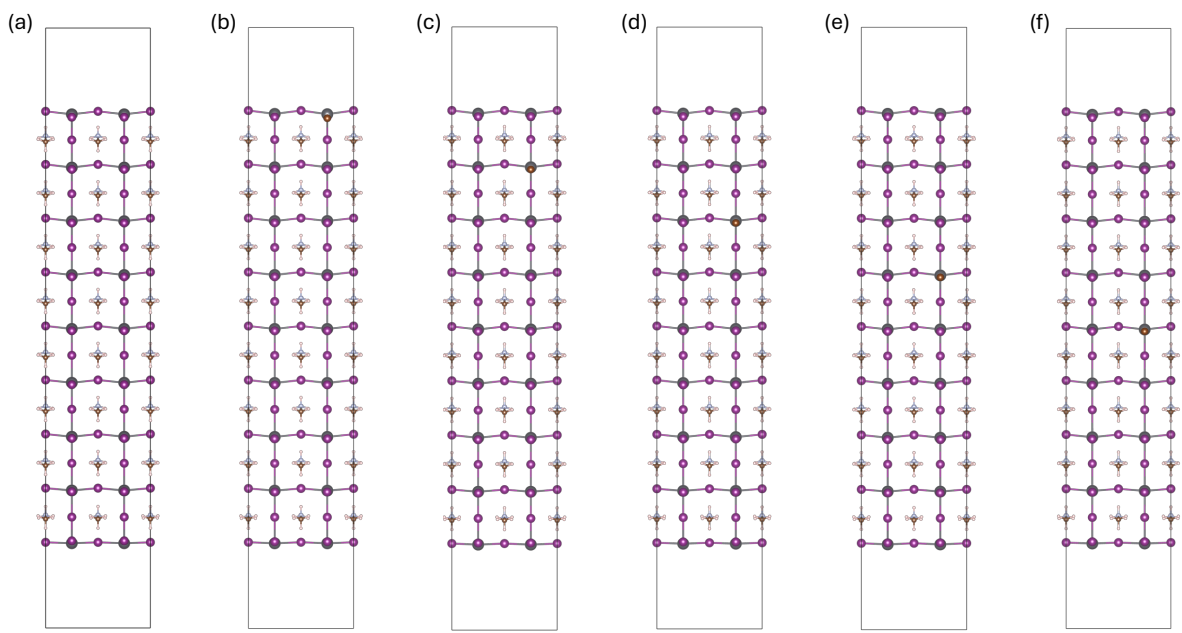

Figure S7: 17-layer MAPbI$_3$ slab with only iodine. It has alternating layers of PbI$_2$ and MAI. 10 Å vacuum is put on both sides of the slab. (a) The pristine slab, (b)Br antisite defect in the first layer, (c)Br antisite defect in the third layer,(d)Br antisite defect in the fifth layer,(e)Br antisite defect in the seventh layer,(f)Br antisite defect in the ninth layer

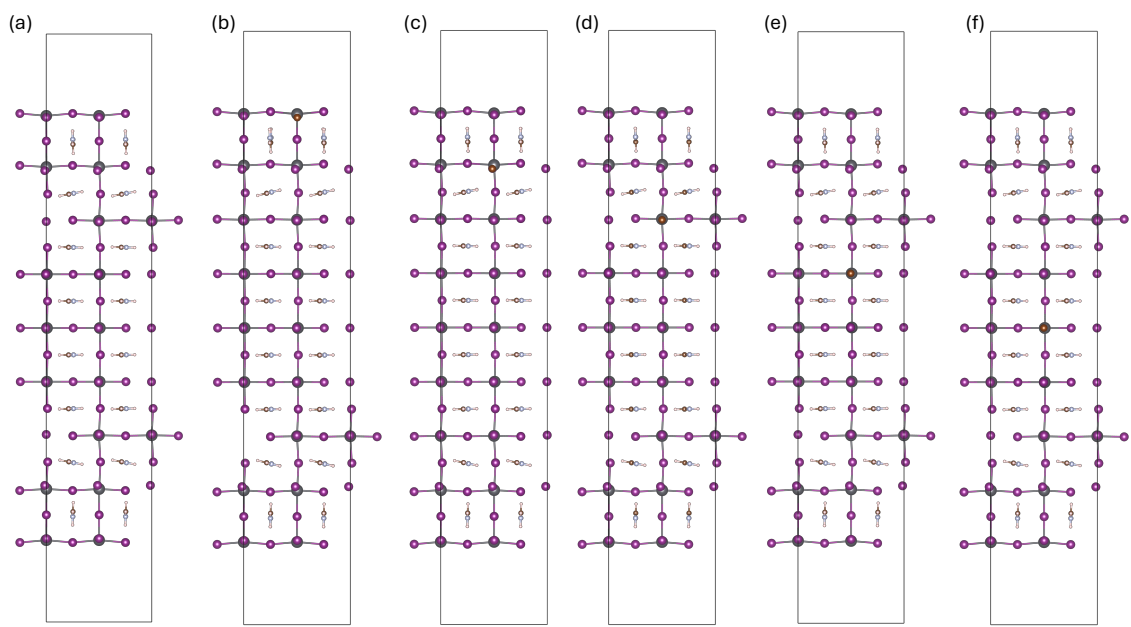

Figure S8: 17-layer FAPbI$_3$ slab with only iodine. It has alternating layers of PbI$_2$ and FAI. 10 Å vacuum is put on both sides of the slab. (a) The pristine slab, (b)Br antisite defect in the first layer, (c)Br antisite defect in the third layer,(d)Br antisite defect in the fifth layer,(e)Br antisite defect in the seventh layer,(f)Br antisite defect in the ninth layer

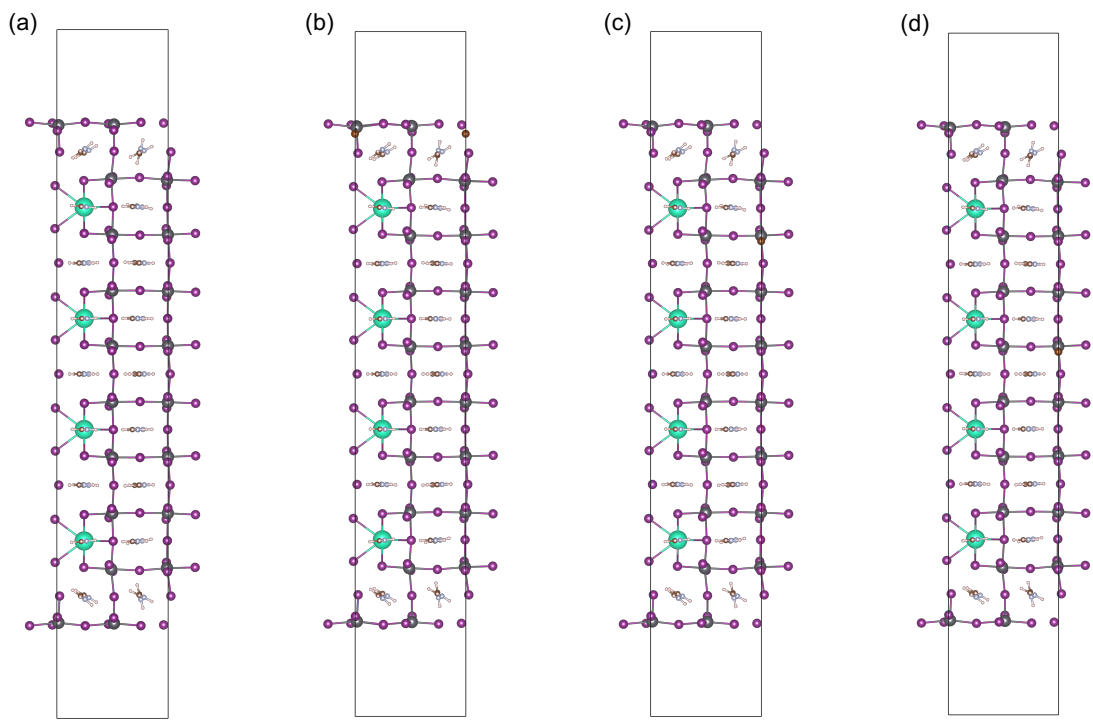

Figure S9: 19-layer $FA_{0.8}Cs_{0.2}PbI_3$ slab with only iodine. 10 Å vacuum is put on both sides of the slab. (a) The pristine slab, (b)Br antisite defect in the first layer, (c)Br antisite defect in the fifth layer,(d)Br antisite defect in the ninth layer

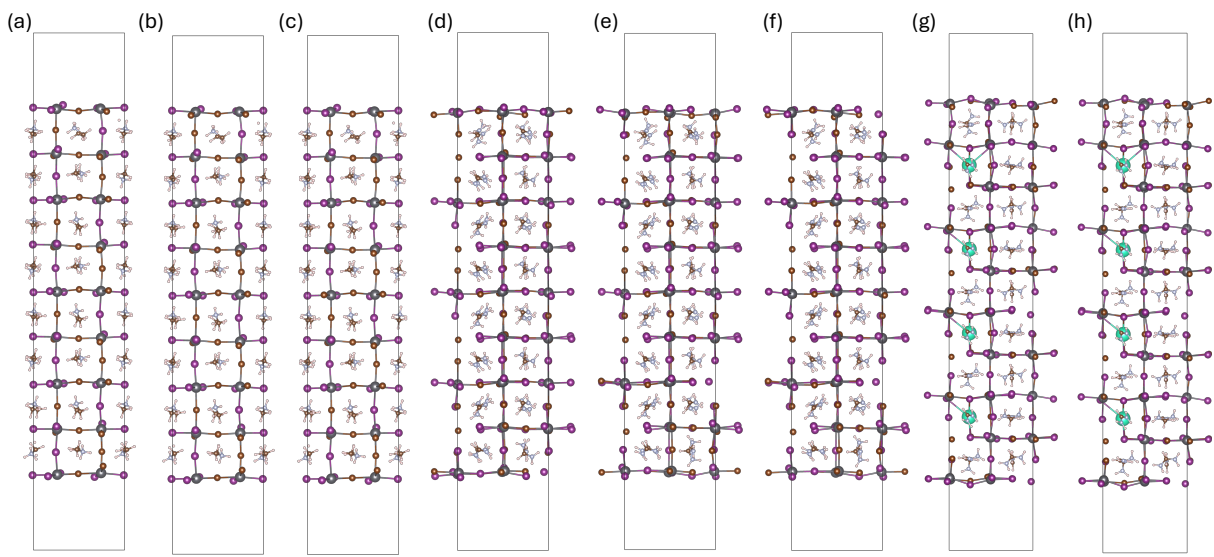

Figure S10: Slabs of various materials with approximately 60% of iodine and 40% of bromine.(a)Pristine MAPb(Br$_{0.4}$I$_{0.6}$)$_3$ slab,(b)MAPb(Br$_{0.4}$I$_{0.6}$)$_3$ slab with a bromine antisite defect in the 1st layer,(c)MAPb(Br$_{0.4}$I$_{0.6}$)$_3$ slab with a bromine antisite defect in the most middle layer,(d)Pristine FAPb(Br$_{0.4}$I$_{0.6}$)$_3$ slab,(e)FAPb(Br$_{0.4}$I$_{0.6}$Br)$_3$ slab with a bromine antisite defect in the 1st layer,(f)FAPb(Br$_{0.4}$I$_{0.6}$)$_3$ slab with a bromine antisite defect in the most middle layer,(g)FA$_{0.8}$Cs$_{0.2}$Pb(Br$_{0.42}$I$_{0.58}$)$_3$ slab with a bromine antisite defect in the 1st layer, (h)FA$_{0.8}$Cs$_{0.2}$Pb(Br$_{0.42}$I$_{0.58}$)$_3$ slab with a bromine antisite defect in the most middle layer

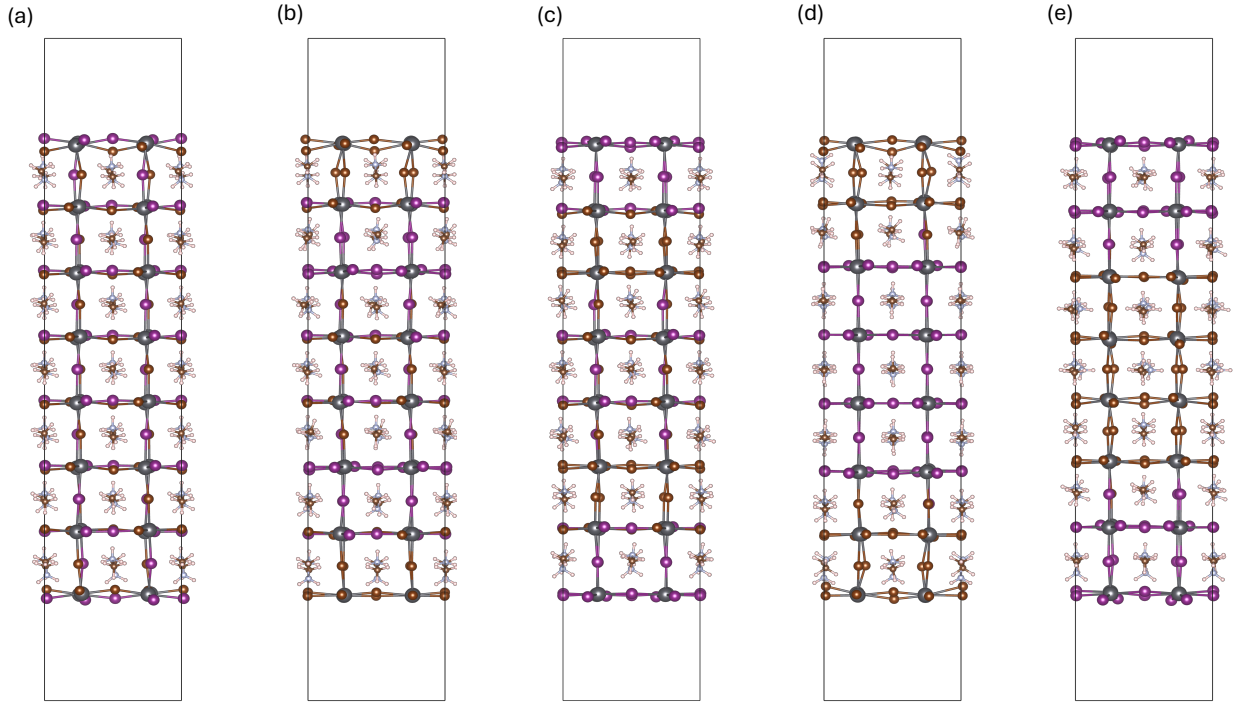

Figure S11: Halide segregation in different MAPb($Br_{0.5}I_{0.5})_3$ 15 layer slabs with 10 Å vacuum.(a)Homogeneous mixture of iodine and bromine. (b)Bromine atoms are near the surfaces where middle 5 layers are kept fixed.(c)Iodine atoms are near the surfaces where middle 5 layers are kept fixed. (d)Bromine atoms are near the surfaces where middle 5 layers are not kept fixed. (e)Iodine atoms are near the surfaces where middle 5 layers are not kept fixed. From first principle calculations, we found that the energy of structure (c) differ from that of (a) by 0.315228845 ev, i.e $\Delta E_c$-$\Delta E_a$, or $\Delta E_{c-a}$ is 0.315228845ev.Similarly we found that $\Delta E_{b-a}$ is -0.4005323ev, $\Delta E_{e-a}$ is 0.287818435ev , $\Delta E_{d-a}$ is -0.77030446. Thus we can infer that as in both cases the energy of the structure with bromine near the surface is the least, bromine atoms will tend to move toward the surface in halide segragation.

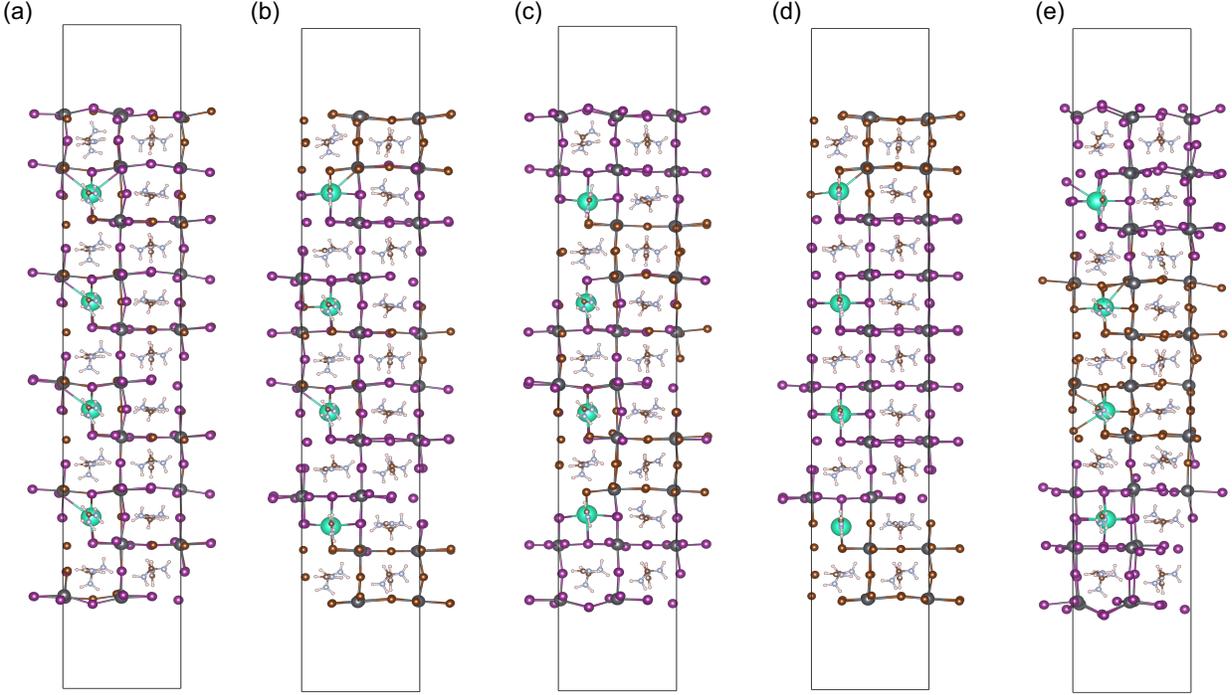

Figure S12: Halide segregation in different $FA_{0.8}Cs_{0.2}Pb(Br_{0.42}I_{0.58})_3$ 19 layer slabs with 10 Å vacuum.(a)Homogeneous mixture of iodine and bromine. (b)Bromine atoms are near the surfaces where middle 5 layers are kept fixed.(c)Iodine atoms are near the surfaces where middle 5 layers are kept fixed. (d)Bromine atoms are near the surfaces where middle 5 layers are not kept fixed. (e)Iodine atoms are near the surfaces where middle 5 layers are not kept fixed. From first principle calculations, we found that the energy of structure (c) differ from that of (a) by 0.04894 ev, i.e $\Delta E_c$-$\Delta E_a$, or $\Delta E_{c-a}$ is 0.04894ev. Similarly we found that $\Delta E_{b-a}$ is -0.21313ev, $\Delta E_{e-a}$ is 1.010809469ev , $\Delta E_{d-a}$ is -0.34996. Thus we can infer that as in both cases the energy of the structure with bromine near the surface is the least, bromine atoms will tend to move toward the surface in halide segragation.

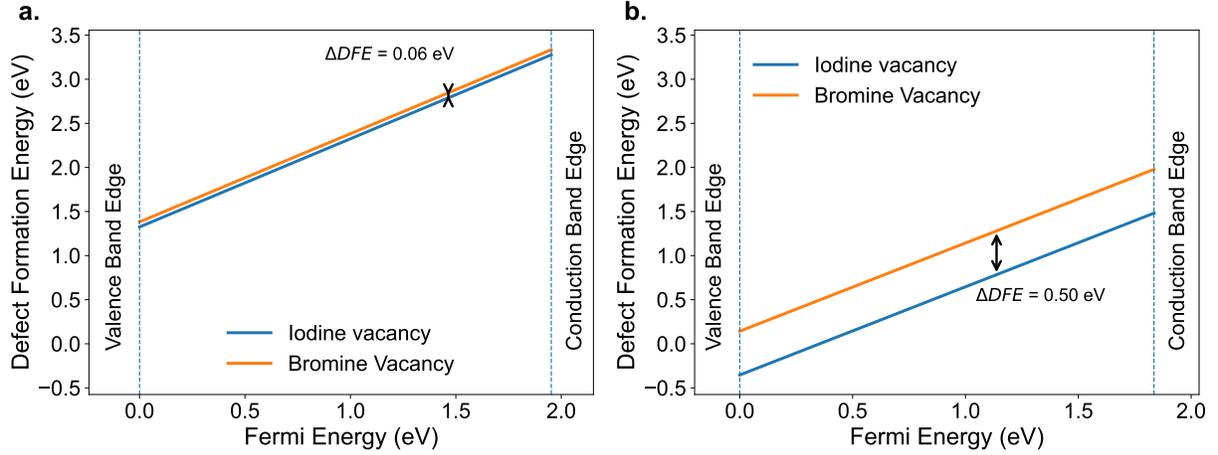

Figure S13: Defect formation energies of iodine ($V_\text{I}^\bullet$) and bromine ($V_\text{Br}^\bullet$) vacancies at the bullk of (a) MAPb(Br$_{0.5}$I$_{0.5}$)$_3$ and (b) FA$_{0.8}$Cs$_{0.2}$Pb(Br$_{0.50}$I$_{0.50}$)$_3$ as a function of Fermi level across the band gap. The valence band maximum (VBM) and conduction band minimum (CBM) are indicated by dashed lines. In both materials, $V_\text{I}^\bullet$ exhibits consistently lower formation energy than $V_\text{Br}^\bullet$ over the full Fermi level range, indicating a stronger thermodynamic preference for iodine vacancy formation.



\end{document}